\documentstyle[preprint,aps,graphicx]{revtex}
\begin{document}
\draft
\title{$\chi$PT Calculations with two-pion Loops for
S-wave $\pi^0$ Production in pp Collision }
\author{ E. Gedalin\thanks{gedal@bgumail.bgu.ac.il},
 A. Moalem\thanks{moalem@bgumail.bgu.ac.il}
 and L. Razdolskaya\thanks{ljuba@bgumail.bgu.ac.il}}
\address
{ Department of Physics, Ben Gurion University, 84105
Beer Sheva, Israel}
\maketitle
\begin{abstract}
The total cross section for the $pp \rightarrow pp \pi^0$ 
reaction at energies close to threshold is calculated within 
the frame of a chiral perturbation theory, taking into account 
tree and one loop diagrams up to chiral order $D=2$. Two-pion
loop contributions dominate $\pi^0$  production at threshold. 
The calculated cross section reproduces data, both scale and 
energy dependence, fairly well.\\

\end{abstract}
\ \\


\newpage

In  recent contributions\cite{park96,cohen96,sato97}, the cross 
section for the $pp \rightarrow pp \pi^0$ reaction at energies
 near threshold was calculated within the frame work of chiral 
 perturbation theory ($\chi$PT). Although $\chi$PT
accounts for all effects such as  unitarity, spontaneously broken
chiral symmetry and offshellness, the calculations in Refs.
\cite{park96,cohen96,sato97} underestimate the cross section 
data by a factor of 3-6. This stands in marked difference
with the results from traditional one-boson exchange (OBE) model
calculations, where contributions from heavy meson exchanges 
seem to resolve the discrepancy between predictions and 
data\cite{lee93,horowitz94,gedalin96}.
Particularly, in a fully covariant OBE model\cite{gedalin96},
 the production amplitude is found to be 
dominated by a  t-pole term, where the pion production occurs on an 
internal meson line at a $\pi\pi\sigma$ - meson vertex. Such 
a mechanism simulates contributions from two-pion exchanges 
and accounts effectively for two-pion loop diagrams.
It is the purpose of the present note to show that the 
failure of $\chi$PT calculations\cite{park96,cohen96,sato97}
to reproduce data may not be due to limitations 
of theory but to inconsistencies
in the way $\chi$PT was applied to this process and, that 
including two-pion loop contributions properly may
resolve the discrepancy between predictions and data.

We carry out $\chi$PT calculations up to chiral order $D=2$, 
taking into account tree and one loop diagrams involving pions 
and nucleons only. The more important of these are depicted in 
Fig. 1. Many other loop diagrams (not shown in Fig. 1) 
contribute very little or/and renormalize the masses and coupling 
constants. The graphs 1a - 1c are the usual impulse and 
rescattering diagrams considered in Refs.\cite{park96,cohen96,sato97}. 
The loop diagrams 1d - 1f correspond to two-pion exchanges 
in t-channels with isoscalar-scalar quantum  numbers. Note that 
the contribution from  graphs 1a-1f can be factorized into a 
pion source , a propagator and an off mass shell amplitude 
for the conversion process $\pi^0p \rightarrow \pi^0p$. 
We shall demonstrate below that this amplitude is strongly 
enhanced due to offshellness. The other graphs 1g and 
1h are contributions specific  to the production process, 
and can not be described in terms of one meson exchanges. 
The latter is a short-range interaction 
mechanism dictated by an order $D=2$ ~$\chi$PT Lagrangian.    

A meson production in NN collisions necessarily involves 
large momentum transfer and two-pion loops like graphs 1d-1g are 
 expected to play an important role.
At threshold the transferred momentum squared
$q^2 = (p_3 - p_1)^2 \approx-Mm$, where $M$ and $m$ are masses 
of the nucleon and meson produced. It is to be demonstrated
that the contribution from diagrams 1d-1f  becomes very
important off the mass shell, and thus providing the enhancement 
required to resolve the discrepancy between previous calculations 
and data.

We use the usual $\chi$PT  pion-nucleon sector heavy-fermion formalism 
(HFF) Lagrangian \cite{park93,bernard94,bernard95}
\begin{equation}
L = L^{(0)}\ +\ L^{(1)}\ +\ L^{(2)}\ \ ,
\label{eq:2}
\end{equation}
where,
\begin{eqnarray}
&&L^{(0)} = \frac {1}{2} [ (\partial_{\mu} {\bf\pi})^2 
- m^2 {\bf \pi}^2 ] - \frac{1}{6F^2} [ {\bf\pi}^2 
(\partial_{\mu} {\bf\pi})^2 - ({\bf\pi}
\cdot \partial_{\mu} {\bf\pi})^2 ] + \frac{m^2}{4!F^2}({\bf\pi}^2 )^2  
+ N^{\dagger}(i v\partial)N + \nonumber\\
&& N^{\dagger}\left\{ -\frac {1}{4F^2} {\bf\tau\cdot\pi} \times 
(v\partial) {\bf\pi}
- \frac {1}{F}g_A S^{\mu} {\bf \tau }\cdot [\partial_{\mu}
{\bf \pi} + \frac {1}{6F^2}({\bf \pi \pi} \cdot \partial_\mu {\bf \pi} - 
\partial_{\mu}{\bf \pi \pi}^2) ] \right\}N;	\label{lagl0}\\
&&L^{(1)} = \frac {1}{2M} (v^{\mu} v^{\nu} - g^{\mu \nu})
\left[ N^{\dagger}\partial_{\mu} \partial_{\nu} N +
\frac {1}{4F^2} (iN^{\dagger} {\bf\tau\cdot \pi} \times \partial_{\mu}
{\bf\pi} \partial_{\nu}N + h.c.)\right] \nonumber\\
&& + \frac {g_A}{2MF}[i N^{\dagger} {\bf\tau } 
(v \partial{\bf \pi})S^{\mu} \partial_{\mu}N + h.c.]	\label{lagl1}
\nonumber \\
&& + \frac {1}{2MF^2} N^{\dagger} [(c'_2 -
\frac {1}{4}g^2_A)(v\partial {\bf \pi})^2 - 
c'_3 (\partial_{\mu}{\bf \pi})^2 
- 2c'_1 m^2 {\bf \pi}^2] N +...~,
\end{eqnarray}

and 
\begin{equation}
L^{(2)} = - \frac {d_1}{2MF}[i N^{\dagger} {\bf \tau } 
(v \partial{\bf \pi})
 S^{\mu} \partial_{\mu}N N^{\dagger}N + h.c.] + ....
\end{equation}
Here $\pi$ and $N$ represent pion and nucleon fields, 
$v$ is the nucleon four velocity, $(v\partial)=
v^{\mu}\partial_{\mu}$, F and $g_A$ are the pion radiative decay 
and axial vector coupling constants.
The  dimensionless low energy coupling constants,
$c'_1 = -1.63$, $ c'_2 = 6.20$ and $ c'_3 = - 9.86$, are determined 
from fitting S-wave $\pi^0$N scattering data \cite{bernard95}.

To consider the relative importance of the various graphs in Fig. 1, we 
apply the  modified power counting scheme of Cohen et al.\cite{cohen96}. 
As already shown in Ref.\cite{cohen96}, the impulse and rescattering 
terms (diagrams 1a-1c)  are of the order $\sim(m/M)^{1/2}F^{-3}$ and  
$\sim(m/M)^{3/2}F^{-3}$, respectively. It is easy to show that the loop 
diagrams are of the same order of magnitude as the impulse term.
Consider for example diagram 1d for which the characteristic momentum
squared is $Q^2 = (p_4 - p_2)^2 \approx -Mm$ (see Fig. 1 for notation). 
This graph depends upon  $\pi N$ as well as 
four pion interaction terms in $L^{(0)}$. From Eqn. 2, the 
$\pi NN$ vertices contribute each a factor  $QF^{-1}$ while 
the four pion vertex
contributes a factor  $Q^2F^{-2}$. In addition there is a 
factor of  $Q^{-2}$  from each of the meson propagators,  $Q^{-1}$ from 
the internal nucleon line, and a factor of  $Q^4(16\pi)^{-1}$ from the
loop integral. Altogether, diagram 1d is of the order 
$\sim ~~Q^2 (16\pi F^5)^{-1}~~ \sim ~~mM (16\pi F^5)^{-1}~~
\sim ~~m (4 F^4)^{-1}$.
Numerically, $m/4F \approx (m/M)^{1/2}$ so that using the same
organizing principle as in Ref.\cite{cohen96}, diagram 1d is 
of the same order of magnitude as the impulse term, what brings 
us to conclude that loop  diagrams should not be disregarded.

We now write the primary production amplitude for the 
$pp\rightarrow pp\pi^0$ reaction in the form
\begin{equation}
M^{(in)}(pp \to pp\pi^0) = M^{(1)}_I + M^{(1)}_R + M^{(1)}_L 
+M^{(2)}_L + M^{(2)}_S~,
\label{mattot}
\end{equation}
where
\begin{eqnarray}
&M^{(1)}_I =  \frac {ig_A}{2F (q^2 - m^2)}\left\{(-) 
\frac {g_A^2}{4 M F^2}{\bf q}^2\right\} {\bf p}{\bf \sigma }_1 + 
   [1 \leftrightarrow 2, 3 \leftrightarrow 4]~, \\
\label{matp}
&M^{(1)}_R = \frac{i g_A}{ 2F (q^2 - m^2)}\left\{
  \frac {1}{MF^2}\left[ (c'_2 + c'_3 - \frac{g^2_A}{4}) mq^0 
  - 2c'_1 m^2\right]\right\}{\bf p}{\bf \sigma }_1 
  + [1 \leftrightarrow 2,3 \leftrightarrow 4]~,\\
&M^{(1)}_L = \frac {ig_A}{2F (q^2 - m^2)} \left\{ 
             \frac {g^2_A}{6F^4} (6mq^0 - 2q^2 
-\frac{5}{2}m^2) B(Q^2)\right\} {\bf p}{\bf  \sigma }_1 
+ [1 \leftrightarrow 2,3 \leftrightarrow 4]~, \\
&M^{(2)}_L = - i \frac{1}{24F^5}g^3_A B(Q^2){\bf p}{\bf \sigma }_1 + 
[1 \leftrightarrow 2, 3 \leftrightarrow 4]~,\\
&M^{(2)}_S = i\frac {d_1 m }{2FM} {\bf p}{\bf \sigma }_1 + 
[1 \leftrightarrow 2, 3 \leftrightarrow 4]~.
\label{mats}
\end{eqnarray}
Here the quantities $M_I^{(1)}$, $M_R^{(1)}$, $M_L^{(1)}$
denote the contributions from the impulse, rescattering and one-loop
diagrams 1d-1f. These are written in a factorized form where the
expressions in the curly brackets represent analogous 
contributions to the conversion process $\pi^0 p \rightarrow \pi^0 p$.
$M_L^{(2)}$ and $M_S^{(2)}$ are the contributions from graphs 1g and 1h.
Our notation  is~: 
$p=(M+{\bf p}^2/2M, {\bf p})$ and $k=(\sqrt {m^2+{\bf k}^2/2M},
 {\bf k})$ stand for the incoming proton and pion produced 
momenta in the overall center of mass (CM) frame; 
$Q = (-{\bf p}^2/2M, {\bf p})$ and 
$q = (-{\bf p}^2/2M, -{\bf p})$ are the transferred momenta.
The bracket $[1 \leftrightarrow 2, 3 \leftrightarrow 4]$ 
represents the contribution from the same diagram with 
the proton momenta $p_1,\ p_3$ interchanged with $p_2,\ p_4$, 
respectively.
The expressions for $M_I^{(1)}$, $M_R^{(1)}$ and $M_S^{(2)}$ are 
identical with those obtained by Cohen et al.\cite{cohen96}. The 
evaluation of the loop contribution, though a bit long and tedious, 
is straightforward and will not be given here. Both, $M_L^{(1)}$ and 
$M_L^{(2)}$ depend on the loop function defined to 
be\cite{park93,bernard95} 
\begin{eqnarray}
B(q^2) = (-3 + 2 {\bf p^2} \frac{d}{dq^2}) B_0(q^2)~,\\
B_0(q^2) = -\frac{1}{16 \pi} \int_0^1 dz  \sqrt{m^2 - q^2 z(1 - z)}~.
\end{eqnarray}

In the calculations to be presented below the values of constants 
and masses are taken to be~:~$F=93~ MeV$, $m=135 ~MeV$, $M=938 ~MeV$ 
and $g_A =1.26$. The short range interaction $d_1$ parameter 
is not determined by chiral symmetry. In order to fix its value
we follow a procedure similar to that applied in Ref.\cite{cohen96},
assuming that the short-range interactions originate from 
$\rho$  and $\omega$ vector meson exchanges as depicted in Fig. 2. 
This leads to
\begin{equation}
d_1 =  \frac {f_{\pi NN} F}{2mM}\left( \frac {g^2_{\rho NN}
 (1 + \kappa)}{m^2_{\rho} + Mm} + \frac {g^2_{\omega NN}}
 {m^2_{\omega}+ Mm}\right )~.
\end{equation}
Here $m_{\rho} = 770 ~MeV$, $m_{\omega} = 782 ~MeV$ are masses 
of the $\rho$  and $\omega$ mesons; $f_{\pi NN}$ the $\pi NN$ 
pseudovector coupling 
constant; $g_{\rho NN}$ and $g_{\omega NN}$ the $\rho NN$ and 
$\omega NN$ vector coupling constants; $\kappa$ the ratio 
of tensor to vector $\rho NN$ coupling constants. With these 
taken from the OBEP set of Machleidt\cite{machleidt89} 
one obtains $d_1 = 1.16 ~fm^3$, a value nearly identical with the 
strength derived in Ref.\cite{cohen96}. 

The various contributions to the production amplitude
are drawn in Fig. 3 
vs $\eta$,  the maximum pion momentum in the overall CM frame. We
confirm the observation of Ref.\cite{cohen96} that the rescattering, 
though enhanced by off shell effects, has an  opposite sign to that
of the impulse term. These two terms interfere destructively, and thus  
reinforcing the importance of the other terms. 
In fact these two together with $M_L^{(2)}$ and  $M_S^{(2)}$, 
the terms from graphs 1g and 1h, cancel to large extent. The 
one-loop term  $M_L^{(1)}$ is significantly more important than any 
of the other contributions. 
This  may well be understood by considering the off mass shell 
behavior of the amplitude for the $\pi ^0 p \to \pi^0 p$ 
conversion process. Using the expression in the curly brackets of 
Eqn. 8, the one loop contribution to the on mass shell
conversion amplitude amounts to $T_L^{(1)} = 3m^3/64\pi F^4 = 0.1 ~fm$,
a value already derived by Bernard et al.\cite{bernard95}.  
Off mass shell at $q= (-m/2, -\sqrt {Mm})$ this term becomes 
rather large 
\begin{equation}
T_L^{(1)} = g_A^2MmB(-Mm(1-m/4M))/3F^4 ~\approx 2.43~fm~.
\end{equation}
Using Eqns. 6-7, the analogous quantities from the impulse and 
rescattering terms are
$T_{I}^{(1)}=-1.21~fm$ and $T_{R}^{(1)} = 0.54~fm$, respectively.
Thus off mass shell two-pion loop contributions dominate the 
conversion process also, with ratios  
$M_I^{(1)} : M_L^{(1)} : M_R^{(1)} 
\approx T_I^{(1)} : T_R^{(1)} : T_L^{(1)}
\approx 5 : 2 : 1 $.

In Fig. 4 we draw predictions for the total cross section of 
the $pp \rightarrow pp \pi^0$ reaction along with the data of 
Refs.\cite{bondar95,meyer92}. Final state interactions (FSI) 
influence the energy dependence as well as the scale of the 
cross section.  We treat FSI in an approximate
way by assuming factorization of the S-wave production amplitude 
into a primary production amplitude, $M^{(in)}$ of Eqn. 5, and
an S-wave FSI factor. For a three body process as in our case, the 
latter is identified\cite{moalem94} with the (on mass shell)
amplitude for $\pi NN \rightarrow \pi NN$ elastic scattering. 
We have documented this approximation in length 
elsewhere\cite{moalem95,gedalin96} and shall skip further 
details here. We stress though that, applying this approximation to 
$pp \rightarrow pp \pi^0$ yield very similar corrections in 
comparison with those obtained with FSI 
between the two charged protons only\cite{sato97,horowitz94,meyer92}. 
The cross section calculated with the full amplitude of Eqn. 5 and 
with FSI corrections is drawn as a solid line in Fig. 4. Without
contributions from loop diagrams the cross section (small dashed
curve) is lower by a factor of $\approx 2.5$. To account for the
initial and final pp interactions Sato et al.\cite{sato97}
have used distorted waves obtained by solving the 
Shr\"{o}dinger equation with NN potential. 
These distorted waves are then used to calculate the matrix
elements of the impulse and rescattering terms. It is very reassuring
that the total cross section they have calculated (the solid
line in their Fig. 4) is very close to our predictions 
(small dashed curve). 
The cross section 
calculated without FSI corrections (long dashed line with the
full amplitude; dot-dashed curve without loop contributions) vary
fast with energy due to phase space factor and does not account
neither for the energy dependence nor for the scale.

In summary we have calculated S-wave pion production in
$pp \rightarrow pp \pi^0$ taking into account tree and one 
loop diagrams up to chiral order D=2. We have found that
loop diagrams contribute significantly to the process.
Dynamically, this means that two-pion exchanges play an essential 
role in the production process.

The calculations presented above can be improved by including 
contributions from other degrees of freedom. For example, excitations 
from the $\Delta$ (1232 MeV) nucleon isobar may well contribute 
to any of the graphs a-g in Fig. 1. In view of the large cancellations 
between the various contributions considered above 
it remains still to be verified that the HFF expansion  converges.
Finally, since contributions from D=2 loop 
diagrams have the same order of magnitude as  those from lower 
order terms then it would be important to ascertain 
convergence of the next  D=3 chiral order diagrams as well.

\vspace{1.5 cm}
{\bf Acknowledgments} This work was supported in part 
by the Israel Ministry Of Absorption.
We are indebted to  Z. Melamed
for assistance in computation.

\newpage
\begin{figure}
\includegraphics[scale=0.80]{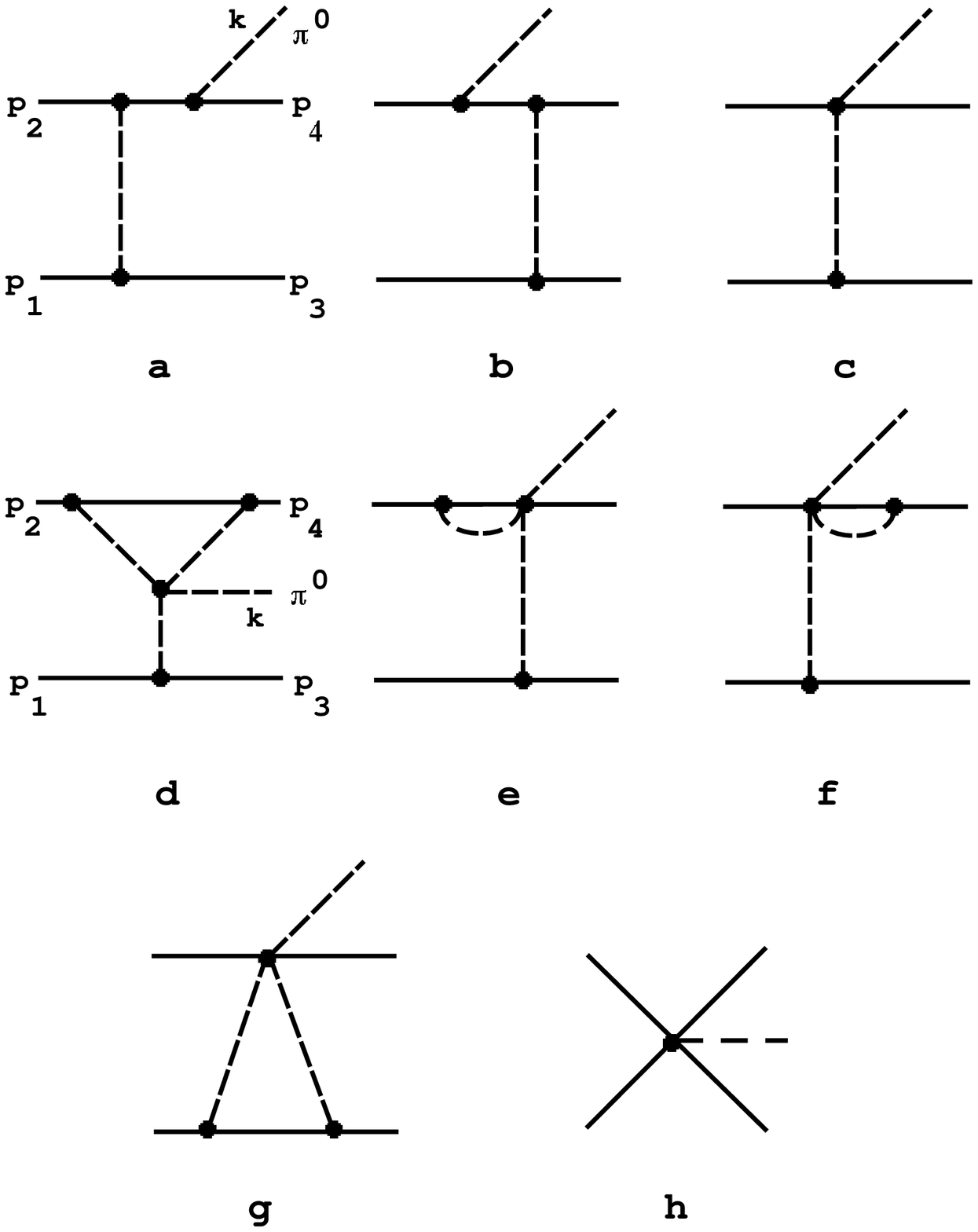}
\caption{Various contributions to the $NN \rightarrow NN \pi^0$ reaction.
In this and following figure a solid line stands for a nucleon and a
dashed line represents a meson. }
\end{figure}
\newpage
\begin{figure}
\includegraphics[scale=0.8]{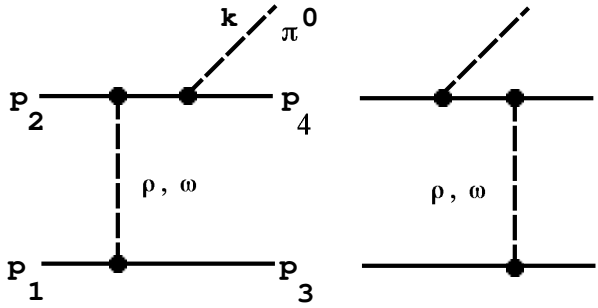} 
\caption{The meson-exchange mechanism used to model the 
short-range interaction.  
Contributions from both $\rho$ and $\omega$ mesons are considered.
}
\end{figure}
\newpage
\begin{figure}
\includegraphics[scale=0.9]{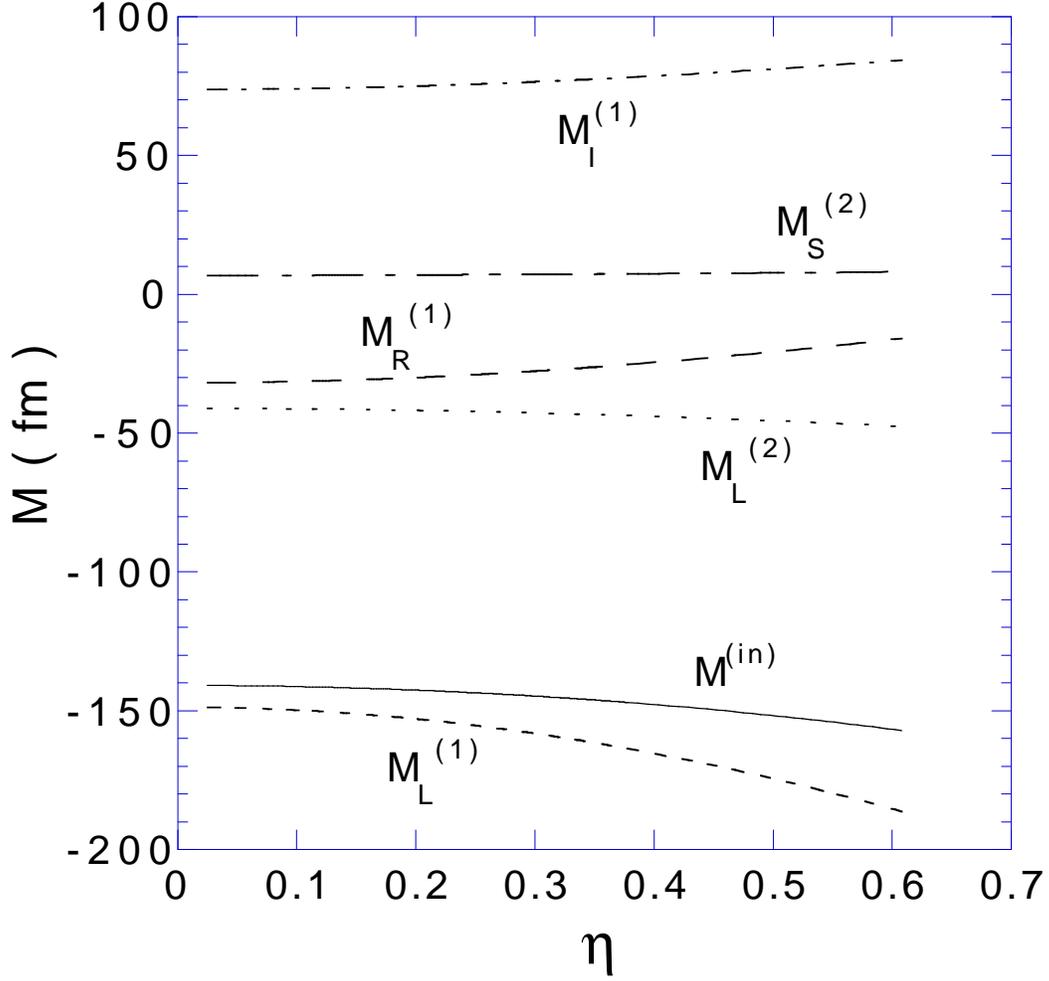}  
\caption{ Various partial amplitudes for the
$pp \rightarrow pp \pi^0$  reaction vs. $\eta$ the maximum pion
momentum in the overall CM frame. The curves labeled, $M_I^{(1)},
M_R^{(1)}, M_S^{(2)}, M_L^{(1)}$ and $M_L^{(2)}$ are due
to the impulse, rescattering, short range interaction and loop
contributions. (see text for details).
}
\end{figure}

\newpage
\begin{figure}
\includegraphics[scale=0.9]{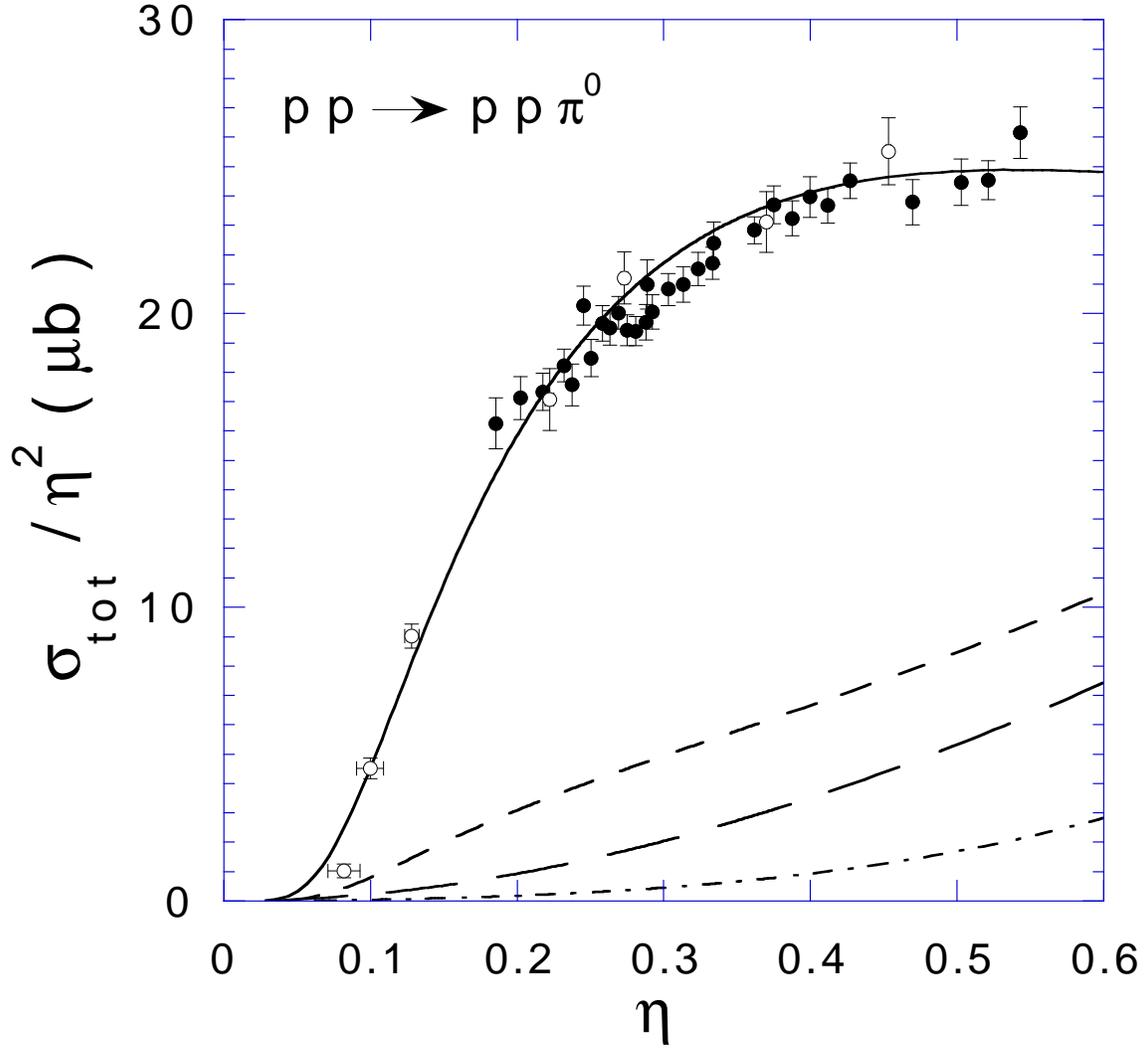}  
\caption{Predictions for the total cross section 
 vs. $\eta$, the maximal pion momentum in the overall CM frame.
Predictions corrected for FSI with (full amplitude of Eqn. 5) and 
without loop contributions are drawn as solid line and small 
dashed curves, respectively. Predictions without FSI corrections 
are drawn as long dashed (full amplitude) and dot-dashed (loop
contributions not included).
}
\end{figure}

\end{document}